\documentstyle[psfig]{mn}

\newcommand{\Dot}[1]{\stackrel{\mbox{.}}{#1}}
\newcommand{\DDot}[1]{\stackrel{\mbox{..}}{#1}}

\begin{document}

\title[Pulsar Braking Indices Revisited]
      {Pulsar Braking Indices Revisited}

\author[Johnston \& Galloway]{Simon~Johnston$^1$ \&
David~Galloway$^{2,1}$\\
$^1$Research Centre for Theoretical Astrophysics, University of Sydney, 
NSW 2006, Australia\\
$^2$School of Mathematics and Statistics, University of Sydney,
NSW 2006, Australia}

\maketitle

\begin{abstract} 
Using the standard equation for the slowdown of a neutron star,
we derive a formula for the braking index
via integration rather than the conventional differentiation. The new
formula negates the need to measure the second time derivative of the
rotation frequency, $\DDot \nu$.
We show that the method gives similar
braking indices for PSR B1509--58 and the Crab pulsar to those already in
the literature. We point out that our method is useful for obtaining the
braking indices of moderate aged pulsars without the need for
long, phase-connected timing solutions. We applied the method to
20 pulsars and discuss the implications of the results.
We find that virtually all the derived braking indices are dominated by
the effects of (unseen) glitches, the recovery from which corrupts the
value of $\Dot \nu$.
However, any real, large, positive braking index has implications for
magnetic field decay and offers support to recent models of pulsar evolution.
\end{abstract}


\section{Introduction}
Neutron stars are powered by rotational kinetic energy and lose
energy by accelerating particle winds and by emitting electromagnetic
radiation at their rotation frequency, $\nu$. The rotation frequency thus
decreases with time and this slowdown is usually described by the relation
\begin{equation}
\Dot \nu = -K\nu^n
\end{equation}
Here, $K$ is a positive constant which depends on the moment of inertia
and the magnetic dipole moment of the neutron star
and $n$ is the braking index.
Conventionally, the braking index is derived by {\it differentiation}
of equation 1, yielding
\begin{equation}
n = \frac{\nu \DDot \nu}{\Dot \nu^2}
\end{equation}
In a highly simplified model in which the spin-down torque arises from
dipole radiation at the rotation frequency, one expects $n=3$.

Only 4 pulsars have had their braking indices measured and all 
have $n<3$. The Crab pulsar has a value $2.509\pm 0.001$
\cite{lps88,lps93}, PSR 1509--58 has a braking index $2.837\pm 0.001$
\cite{kms+94}, PSR 0540--69 has $n=2.04\pm 0.02$ \cite{mp89,ndl+90,gfo92}
and, finally, the Vela pulsar has the low value $1.4\pm0.2$ \cite{lpsc96}.
Melatos (1997)\nocite{mel97} has shown that a modification of the
simple model, which involves treating the neutron star and the
inner magnetosphere as one entity, allows him to derive values of
braking index very close to those observed (except for the Vela pulsar).

Braking indices are very difficult to measure in all the other pulsars.
For a typical `old' pulsar with $\nu=1$~Hz, $\Dot \nu=10^{-15}$~Hz/s,
the expected $\DDot \nu$ from equation 2 is only $\sim 10^{-30}$~Hz/s$^2$,
much too small to measure even over hundreds of years of timing - the
second derivative only contributes one extra phase rotation every
600 yr!
There are a number of pulsars with ages $\sim$20~kyr for which one
might expect to be able to measure $n$. However, two different effects tend
to dominate the value of $\DDot \nu$ over that expected from spin-down alone.
First, these young pulsars glitch often \cite{sl96}. These glitches lead
to discontinuities in both $\nu$ and $\Dot \nu$ making it very difficult
to phase connect (i.e. count the exact number of rotations of the pulsar)
over the glitch. Furthermore, the recovery from a glitch
can last many hundreds of days and the measurement of $\Dot \nu$
reflects the recovery rather than the intrinsic spin-down.
Finally, young pulsars have large random variations in arrival times
known as `timing noise'. Cordes \& Helfand (1980)\nocite{ch80} recognised
that the timing noise dominates over the intrinsic
$\DDot \nu$ by a factor $\sim$100 in these pulsars and that many early
published values of braking indices, based on $\DDot \nu$, were spurious.

\section{Braking Index by Integration}
Instead of differentiating equation 1, we integrate from a time $t$
to $t+T$ to obtain
\begin{equation}
\left[\frac{\nu^{1-n}}{1-n}\right]^{\nu_2}_{\nu_1} = -KT = \frac{\Dot \nu_{1}}{\nu^{n}_{1}} \, T
\end{equation}
from which follows
\begin{equation}
\frac{\nu^{1-n}_{2} - \nu^{1-n}_{1}}{1-n} = \frac{\Dot \nu_1 T}{\nu^{n}_{1}}
\end{equation}
Hence
\begin{equation}
1-n = \frac{1}{\Dot \nu_1 T} \left[ \nu_2 \left(\frac{\nu_1}{\nu_2}\right)^n - \nu_1 \right]
\end{equation}
But, $(\nu_1/\nu_2)^n = \Dot \nu_1/\Dot \nu_2$ and so
\begin{equation}
\label{new}
n = 1 + \frac{\nu_1\Dot \nu_2 - \nu_2\Dot \nu_1}{\Dot \nu_1 \Dot \nu_2 T}
\end{equation}

This allows the braking index to be computed without the need to measure
$\DDot \nu$. The advantage of this method is that, in principle,
$\nu$ and $\Dot \nu$ can be measured over a short interval of time and
then re-measured 20 yr later without the need for a phase connected
solution over the whole 20 yr time span.

\subsection{Error analysis}
We can rewrite equation \ref{new} as
\begin{equation}
n = 1 + \frac{1}{T} \left( \frac{\nu_1}{\Dot \nu_1} - \frac{\nu_2}{\Dot \nu_2} \right)
\end{equation}
Thus, for $n\sim 3$, the value of the expression inside the brackets must
be $\sim 2T$. Typically, the fractional error in $\Dot \nu$ is much larger
than that in $\nu$.  Let $\Dot e_1$ and $\Dot e_2$ be the error in
$\Dot \nu_1$ and $\Dot \nu_2$. The absolute error, $E$, on the braking index
is thus given by
\begin{equation}
\label{error}
%E \simeq \frac{2\,\,\,\nu}{T \Dot \nu^2}\,\,\,(\Dot e_1 + \Dot e_2)
E \simeq \frac{\nu}{T \Dot \nu^2}\,\,\,\sqrt{\Dot e_1^2 + \Dot e_2^2}
\end{equation}

Taking a young pulsar such as the Crab as an example,
the parameters for which are
$\nu\sim~30$~Hz, $\Dot \nu~\sim~-3.8\times 10^{-10}$~Hz/s and
$\Dot e~\sim~10^{-15}$~Hz/s over one month \cite{lps88}. If measurements are
made $\sim$1~yr apart then the error in the braking index is $\sim$0.01.

For older pulsars, those near 100~kyr, one can obtain a reasonably
low error on the braking index, provided the time span is large enough.
For example, 
PSR B0540+23 was discovered more than 25 yr ago. Its parameters
are $\nu\sim~4$~Hz, $\Dot \nu~\sim~-2.5\times 10^{-13}$~Hz/s and
$\Dot e~\sim~10^{-18}$~Hz/s \cite{antt94}. Equation \ref{error} yields an 
estimated error in the braking index of $\sim$0.1, accurate enough to
distinguish between various slow-down models.

\section{Discussion}
\subsection{Pulsars with known braking index}
For the Crab pulsar,
Lyne et al. (1988)\nocite{lps88} have tabulated $\nu$, $\Dot \nu$
and their associated errors at monthly intervals over a period of five
years. In order to determine the braking index, they first compute 
$\DDot \nu$ by performing a straight line fit to the $\Dot \nu$ versus
time graph. They derive a braking index of $2.509\pm 0.001$ based on the
value of $\DDot \nu$ over this interval.
We can use their tabulated values and derive a braking index from
equation \ref{new} above, without the need to compute $\DDot \nu$
explicitly. We can do this for every pair of values
in the table (i.e. 2278 pairs in total). The results are displayed
in Fig. 1.

It can clearly be seen that for the shortest intervals, both measurement
error and short-term timing noise dominate the value of the braking
index. For time intervals greater than $\sim$500 days, the braking
index is stable near 2.5. This is consistent with the $\sim$600 days
periodicity seen in the timing noise by Lyne et al. (1988).
The weighted average of all the measurements is 2.502 (however, the
braking indices are not truly independent); the braking
index measurement from the largest time interval is $2.516\pm 0.003$.
Both these measurements are consistent with the Lyne et al. (1988) result.
It is not possible to compute a braking index for the Crab over a longer 
period of time. Lyne et al. (1993)\nocite{lps93} have shown that the
glitches permanently alter the value of $\Dot \nu$; however the inter-glitch
value of the braking index remains roughly constant.

Kaspi et al. (1994)\nocite{kms+94} published a braking index 
of $2.837\pm 0.001$ for PSR B1509--58
based on a phase-connected solution across 11 yr of timing data.
This pulsar has not been observed to glitch in that interval, and the
timing noise is surprisingly low for such a young pulsar.
We applied our method by finding a local fit to $\nu$ and $\Dot \nu$
over a period of 3 months in 1993 and 1997. Over this interval
of 1700 days we find $n=2.80\pm 0.03$ consistent with the Kaspi et al. (1994)
result but without the need for a phase connected solution.

\begin{figure}
\psfig{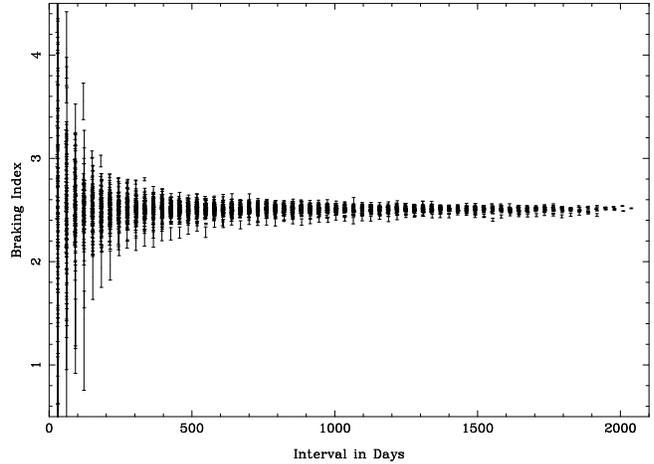}
\caption{Braking index for the Crab pulsar as a function of time over
the period 1982-1987. The braking index is computed using equation \ref{new}
above and parameters for $\nu$ and $\Dot \nu$ from Lyne et al. (1988).
2278 data points are shown.}
\end{figure}

\subsection{Young, Vela-like pulsars}
There are 21 pulsars with $|\Dot \nu| > 10^{-12}$~Hz/s.
Of these, 4 have their braking
indices measured (see above), and 11 of the remaining 17 have had one
or more glitches and are expected to glitch every $\sim$10 yr or so (see
e.g. Shemar \& Lyne 1996)\nocite{sl96}.
For the Vela pulsar, for example, the
intrinsic $\DDot \nu$ is completely dominated by the recovery of
$\Dot \nu$ following the regular glitches. Lyne et al. (1996)\nocite{lpsc96}
solved this problem by taking $\Dot \nu$ at a fixed interval after each
glitch and deriving $\DDot \nu$ and hence the braking index from
the changing $\Dot \nu$. Presumably, if such a method is valid, one
could apply equation \ref{new} over the largest time interval available
(25 yr) to obtain an error on the braking index of $\sim$0.1.

The six pulsars which are not known to have glitched are PSRs B1951+32,
B1853+01, B1930+22,
B1643--43, B0906--49 and J0631+1036. Taking PSR B1951+32 as an example,
Foster et al. (1994)\nocite{flsb94}
have shown that this pulsar suffers from excessive timing noise.
By fitting for higher orders in $\nu$ they obtain a value of
$\DDot \nu$ of $2.7\times 10^{-22}$~Hz/s$^2$.
This is about a factor $\sim$50 higher than that
expected from the simple spin-down, making it virtually impossible
to compute a braking index.
PSR B0906--49 was discovered 12 yr ago, and has not glitched
since discovery. Using recent timing data, the value of $\DDot \nu$
from timing noise is $\sim 1.8\times 10^{-23}$~Hz/s$^2$, more
than 30 times the expected spin-down value.
Thus, timing noise is dominating the value of $\Dot \nu$
in the literature \cite{dmd+88} and a braking index cannot be computed.
In particular the formal fitting errors given by {\sc TEMPO} (the least-squares
pulsar timing package) are
seriously underestimating the systematic effects of the timing noise.

For these young pulsars, the new method
suffers from similar problems to the standard
calculation of the braking index. In Vela-like pulsars glitches dominate
the overall spin-down behaviour, masking the true spin-down $\Dot \nu$.
In slightly older pulsars, glitches are less frequent; however, the
true $\DDot \nu$ is hidden in large timing noise residuals.

\subsection{Pulsars with moderate $\Dot \nu$}
As described above, it should be possible in principle to obtain
an error on the braking index as low as $\sim$0.2 for pulsars of
age $\sim 10^5$~yr discovered more than 20~yr ago (especially those
with low timing noise).
We select pulsars in the following way. We construct an `expected' value
of $\DDot \nu$ for every pulsar, assuming a braking index of 3 and
select all those with $\DDot \nu > 10^{-27}$~Hz/s$^2$.
We reject all those pulsars which are known to have glitched in the
past and the 21 pulsars described in 3.2 above.
We then searched the literature for at least two timing solutions
for the remaining pulsars 
and with a $\Dot \nu$ of sufficient accuracy to allow the error
on the computed braking index to be less than 20.
We discovered, however, that in some publications, the epoch of $\nu$ and
$\Dot \nu$ corresponded not to the middle of the data span, but rather to the
start. If there is a significant value of $\DDot \nu$ in the data span
which is unaccounted for (either related to timing noise or to the
intrinsic spin-down), this leads to an error in the quoted $\Dot \nu$ of
$\sim \frac{1}{2}\DDot \nu T$ which, for large $T$, can be significantly 
larger than the formal fitting error. This led to the rejection of 
a further five pulsars.

Table 1 gives the braking index for 20 pulsars which survive the above
selection criteria.  Column 1 of the table gives the pulsar name.
Columns 2 and 3 give the pulsar's rotation frequency
and its first derivative, and column 4 the time interval between the epochs
of measurement. Column 5 gives the braking index and associated error
according to equations \ref{new} and \ref{error}. The final column gives
the references for the timing solutions.
Note that for PSR B0656+14 the value of $\Dot \nu$ given in
Ashworth \& Lyne (1981)\nocite{al81} is clearly in error. Hence, the time
interval between timing solutions is rather shorter than it might have been.

The values of the braking indices listed in the table raise the question
as to how realistic the quoted errors are. Typically (but not always) in
timing solutions the errors are twice the formal standard error given by
the {\sc TEMPO} least-squares timing package. However, the errors
are computed assuming that the each TOA has uncorrelated residuals (i.e. white
noise), which is clearly not the case in pulsars
with significant (red) timing noise. Thus the error on $\Dot \nu$ is likely
to be larger than that given from the fit alone. Also, as described above,
not fixing the epoch in the middle of the data span leads to errors
in the value of $\Dot \nu$.
\begin{center}
\begin{tabular}{rlcrccc} \hline
& PSR B & $\nu$ & $\Dot \nu\times 10^{-15}$  & Interval & $n$ & Refs \\
&       & (Hz)  & \multicolumn{1}{c}{(Hz/s)} &  (days)  \\
\hline
1  & 0114+58  & 9.86 & --562.3 & 2271.1 & --9.6$\pm$1.5  & 1,2 \\
2  & 0136+57  & 3.67 & --144.2 & 4492.0 & --81$\pm$4.7   & 5,2 \\
3  & 0154+61  & 0.43 & --34.2  & 4336.5 & 28$\pm$14      & 5,2 \\
4  & 0540+23  & 4.06 & --255.0 & 5543.5 & 11.1$\pm$8.6   & 8,2 \\
   &          &      &         & 5990.5 & 11.81$\pm$0.12 & 3,2 \\
5  & 0611+22  & 2.99 & --530.8 & 5541.5 & 20.1$\pm$1.1   & 3,2 \\
6  & 0656+14  & 2.60 & --371.5 & 2163.3 & 14.7$\pm$1.4   & 1,2 \\
7  & 0740--28 & 6.00 & --604.9 & 4245.2 & 17.7$\pm$1.4   & 9,7\\
   &          &      &         & 5827.2 & 25.6$\pm$0.8   & 9,2\\
8  & 0919+06  & 2.32 & --74.0  & 4521.7 & 28.9$\pm$4.1   & 5,2\\
9  & 1221--63 & 4.62 & --105.7 & 6661.3 & 18.7$\pm$12.3  & 6,10\\
10 & 1356--60 & 7.84 & --389.9 & 7050.3 & 6.3$\pm$6.8    & 6,10\\
11 & 1719--37 & 4.23 & --194.6 & 4824.0 & --183$\pm$10   & 6,2 \\
12 & 1742--30 & 2.72 & --79.0  & 1581.0 & --132$\pm$5    & 7,2 \\
13 & 1829--08 & 1.54 & --151.4 & 1541.0 & 2.5$\pm$0.9    & 4,2 \\
14 & 1907+10  & 3.52 & --32.8  & 5842.5 & 24$\pm$17      & 3,2 \\
15 & 1914+09  & 3.70 & --34.5  & 5556.6 & --15$\pm$16    & 3,2 \\
16 & 1915+13  & 5.14 & --190.0 & 6080.5 & 36.08$\pm$0.48 & 3,2 \\
17 & 2000+32  & 1.44 & --216.7 & 1381.0 & --226$\pm$4.5  & 4,2 \\
18 & 2002+31  & 0.47 & --16.7  & 6076.5 & 23.3$\pm$1.0   & 3,2 \\
19 & 2148+52  & 3.01 & --91.2  & 2307.2 & 49.6$\pm$3.5   & 1,2 \\
20 & 2334+61  & 2.02 & --776.2 & 2347.1 & 8.60$\pm$0.13  & 1,2 \\
\hline
\end{tabular}
\end{center}
[Table 1 references: 1-Dewey et al. (1988), 2-Arzoumanian et al. (1994),
3-Gullahorn \& Rankin (1978), 4-Clifton et al. (1992),
5-Backus, Taylor \& Damashek (1982), 6-Newton, Manchester \& Cooke (1981),
7-Siegman, Manchester \& Durdin (1993), 8-Helfand et al. (1980),
9-Manchester et al. (1983), 10-Parkes timing programme.]
\nocite{gr78,antt94,dtms88,clj+92,btd82,nmc81,smd93,htbc80,mnhg83}

Figure 2 displays the braking index for these 20 pulsars with error bars
a factor of 5 larger than in the table. We believe these error bars
are conservative and more accurately reflect the contributions from the 
underlying timing noise to the intrinsic value of $\Dot \nu$.
From the figure, 14 of the 20 pulsars have significant values of
braking index. The eight pulsars with the smallest error bars all have
positive braking indices (PSRs B0540+23, B0611+22, B0656+14, B0740--28,
B1915+13, B2002+31, B2148+52 and B2334+61).
The four pulsars with large negative braking indices are
PSRs B0136+57, B1719--37, B1742--30, B2000+32 and B2255+58.
\begin{figure}
\psfig{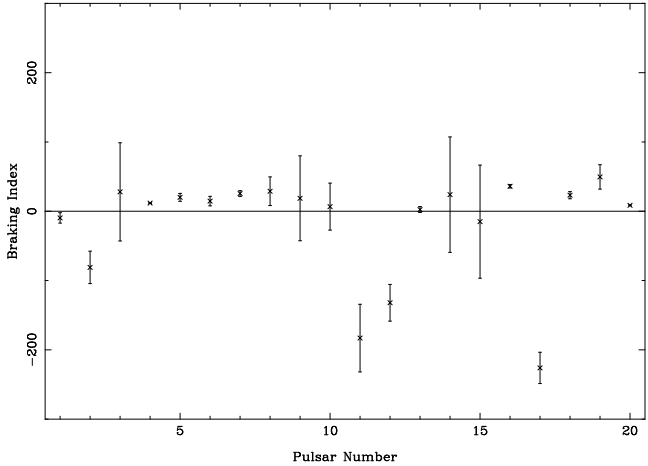}
\caption{Braking index for 20 pulsars from Table 1. The error bars are 
a factor of 5 larger than in the table.}
\end{figure}

\subsection{Implications}
The presence of glitches in pulsars can lead to spurious values of the
braking index and we surmise that (unseen) glitches are the main
cause for $n\neq 3$. If we assume that glitches cause $\Delta\Dot\nu / \Dot\nu$
to be positive in all cases and that the value of $\Dot\nu$ recovers
to nearly its original value, then large negative braking indices can
be caused by glitches between timing solutions and positive braking
indices arise when the glitch occurs before the epoch of the first timing 
solution.  Alpar \& Baykal (1994)\nocite{ab94}, in a statistical survey of the
frequency of pulsar glitches, showed that the number of glitches, $g$, in a
sample of pulsars is given by
\begin{equation}
g = \left( \frac{\delta \nu}{\nu}\right)^{-1} \,\,\,
\sum\nolimits T \,\, \frac{\Dot \nu}{\nu}
\end{equation}
where $T$ is the time interval over which the pulsar is monitored and
$\delta \nu/\nu$ is $1.74\times 10^{-4}$ \cite{ab94}.

Using the values of $\nu$, $\Dot \nu$ and $T$ from the table, we
derive $g=2.8$, i.e. one expects there to have been glitches in $\sim$3
pulsars in between their timing solutions. We have four pulsars
with large negative braking indices which is consistent with this estimate.
The values of $\Delta\Dot\nu / \Dot\nu$ for these four pulsars
are 1.3, 3.6, 0.5, 4 and 0.6 $\times 10^{-3}$ respectively, values
typical of glitches \cite{sl96}. It is thus possible that these pulsars
have all glitched between the epochs of the observations. However, none
of the 4 appear to have suffered any change in $\nu$, i.e. any step
change at the time of the glitch would have been recovered in the
intervening time period.
Alternative explanations are hard to find - for PSR B2000+32, for example,
one of the two timing solutions would have to be in error by more than
100$\sigma$ to obtain a `real' braking index of 3.

For the positive values of braking index, it is harder to compute the
number of glitches expected {\it before} the timing solutions. Both
the recovery time and the form of the recovery (linear or
exponential) for the change in $\Dot \nu$ are unclear.
Shemar \& Lyne (1996) estimate that the recovery time
is greater than 3 yr for the exponential case or tens
of years if the recovery is linear. Indeed, in some pulsars, a permanent
alteration in the value of $\Dot \nu$ is observed. If we let $T=25$~yr,
then we would expect 6.4 glitches prior to the first
timing solution for these 20 pulsars. We have 8 positive braking indices,
broadly in line with this estimate (and we also note that the 2 pulsars
with the highest glitch probability are PSRs B2334+61 and B0611+22, both
of which have positive braking indices).

A number of other factors may contribute to $n\neq 3$.
It is possible that the different methods of obtaining the pulsar 
position could affect the results (however, for PSR B0540+23, for example, 
the positions are identical within the errors).
Another possibility is that the different fitting software,
clock corrections, ephemeris changes etc can affect the result but
it is difficult to believe this could have an effect at the
500$\sigma$ level. These possibilities could be overcome by recovering
the original TOAs from the 1970s and reducing them with the same
software as the 1990s data.

The glitch interpretation put forward above, is statistical in nature.
It may be that in one or more cases the braking index is indeed
significantly larger than 3. What are the implications of this?
Blandford \& Romani (1988)\nocite{br88} showed that if the value of $K$
in equation 1 is time variable then one can write
\begin{equation}
\frac{\Dot K \nu}{K \Dot\nu} = n_{\rm obs} - n.
\end{equation}
$K/\Dot K$ then gives the timescale for variation, if we assume that $n=3$.
For the cases in which $n_{\rm obs} > 3$, $\Dot K < 0$, and the timescale
is surprisingly constrained in the range $\sim$14 to $\sim$56 kyr.
$\Dot K < 0$ can arise from magnetic field decay,
alignment of the spin and magnetic axis,
or an increase in the moment of inertia. Evidence for and against
magnetic field decay and axis alignment have raged in the literature
for 25 yr without any clear consensus emerging. Recently, Chen,
Ruderman \& Zhu (1998)\nocite{crz98} have developed a model in which
the magnetic field increases by a factor of $\sim$10 over the first 10 kyr
of a pulsar's life (which may explain the {\it low} braking index of
the Vela pulsar) and then decreases thereafter. In their model, they expect
a braking index of $\geq$5 for middle aged pulsars before it settles in
older pulsars back to 3. If any of the above braking indices are thus real,
this implies a magnetic field decaying by a factor of $\sim$100 in
$\sim$200 kyr, lending support to the Chen et al. (1998) model.

In any case, one is forced to conclude that the initial starting
premise that the spin-down of a pulsar can be described as a simple
power law in $\nu$ is highly questionable. In virtually all young pulsars,
and in most older ones, complex glitch behaviour appears to be
dominating the spin-down behaviour. This glitch behaviour,
coupled with changes in the magnetic field strength over time, indicates
that the entire concept of a `braking index' must be treated with some caution.

Verification of the glitch hypothesis or the presence of a true braking
index could be obtained by deriving
additional timing solutions for these pulsars now. This would give a 
7 yr span back to the Arzoumanian et al. (1994) data and, in some
cases, more than 20 yr back to earlier data. One might then expect
to observe the exponential recovery of $\Dot \nu$ in the case of a glitch.

\section{Conclusions}
We have derived a new method for computing pulsar braking indices
based on integration of the standard slow-down equation.
We show that the method replicates the known values for both the
Crab pulsar and PSR B1509--58.
The new method conveys no advantage over the previous method for
pulsars with regular glitches and/or large timing noise.
The advantage of equation \ref{new} is that for pulsars with ages
$\sim 10^5$ yr where both the timing noise and the glitch frequency
are relatively low, it is not necessary to obtain a phase connected
solution across 25~yr or more in order to obtain a braking index.
Rather, local fits over 1-2 yr at large intervals may be sufficient
to yield the braking index to an error of $\sim$0.1.

We computed braking indices for 20 pulsars based on timing solutions
available in the literature.
Four pulsars have large negative braking indices
which we believe have been caused by glitches occurring in between the
epochs of the timing solutions. A further eight pulsars
have moderate positive braking indices which appear to be robust to the 
effects of timing noise but are possibly due to glitches which
occurred before the start of the timing observations.
If any of the large, positive braking indices are real, it may provide
evidence for magnetic field decay in moderate aged pulsars.
However, the entire concept of a smooth pulsar spin-down and of a 
constant braking index must be treated with some caution.

\section*{Acknowledgments}
We thank the ATNF pulsar group for providing TOAs for PSRs B1221--63,
B1356--60 and B1509--58.
We also thank L. Ball, R. Manchester and M. Wardle for 
comments on the manuscript.
We acknowledge use of the {\sc TEMPO} software package for pulsar timing.


\begin{thebibliography}{{Gouiffes, Finley \& \"{O}gelman }{1992}}
 
\bibitem[\protect\citename{Alpar \& Baykal }{1994}]{ab94}
Alpar~M.~A., Baykal~A., 1994, MNRAS, 269, 849

\bibitem[\protect\citename{Arzoumanian {\rm et~al. }}{1994}]{antt94}
Arzoumanian~Z., Nice~D.~J., Taylor~J.~H., Thorsett~S.~E., 1994, ApJ, 422, 671

\bibitem[\protect\citename{Ashworth \& Lyne }{1981}]{al81}
Ashworth~M., Lyne~A.~G., 1981, MNRAS, 195, 517

\bibitem[\protect\citename{Backus, Taylor \& Damashek }{1982}]{btd82}
Backus~P.~R., Taylor~J.~H., Damashek~M., 1982, ApJ, 255, L63

\bibitem[\protect\citename{Blandford \& Romani }{1988}]{br88}
Blandford~R.~D., Romani~R.~W., 1988, MNRAS, 234, 57P

\bibitem[\protect\citename{Chen, Ruderman \& Zhu }{1998}]{crz98}
Chen~K., Ruderman~M., Zhu~T., 1998, ApJ, 493, 397

\bibitem[\protect\citename{Clifton {\rm et~al. }}{1992}]{clj+92}
Clifton~T.~R., Lyne~A.~G., Jones~A.~W., McKenna~J., Ashworth~M., 1992, MNRAS,
  254, 177

\bibitem[\protect\citename{Cordes \& Helfand }{1980}]{ch80}
Cordes~J.~M., Helfand~D.~J., 1980, ApJ, 239, 640

\bibitem[\protect\citename{D'Amico {\rm et~al. }}{1988}]{dmd+88}
D'Amico~N., Manchester~R.~N., Durdin~J.~M., Stokes~G.~H., Stinebring~D.~R.,
  Taylor~J.~H., Brissenden~R. J.~V., 1988, MNRAS, 234, 437

\bibitem[\protect\citename{Dewey {\rm et~al. }}{1988}]{dtms88}
Dewey~R.~J., Taylor~J.~H., Maguire~C.~M., Stokes~G.~H., 1988, ApJ, 332, 762

\bibitem[\protect\citename{Foster {\rm et~al. }}{1994}]{flsb94}
Foster~R.~S., Lyne~A.~G., Shemar~S.~L., Backer~D.~C., 1994, AJ, 108, 175

\bibitem[\protect\citename{Gouiffes, Finley \& \"{O}gelman }{1992}]{gfo92}
Gouiffes~C., Finley~J.~P., \"{O}gelman~H., 1992, ApJ, 394, 581

\bibitem[\protect\citename{Gullahorn \& Rankin }{1978}]{gr78}
Gullahorn~G.~E., Rankin~J.~M., 1978, AJ, 83, 1219

\bibitem[\protect\citename{Helfand {\rm et~al. }}{1980}]{htbc80}
Helfand~D.~J., Taylor~J.~H., Backus~P.~R., Cordes~J.~M., 1980, ApJ, 237, 206

\bibitem[\protect\citename{Kaspi {\rm et~al. }}{1994}]{kms+94}
Kaspi~V.~M., Manchester~R.~N., Siegman~B., Johnston~S., Lyne~A.~G., 1994, ApJ,
  422, L83

\bibitem[\protect\citename{Lyne, Pritchard \& Smith }{1988}]{lps88}
Lyne~A.~G., Pritchard~R.~S., Smith~F.~G., 1988, MNRAS, 233, 667

\bibitem[\protect\citename{Lyne, Pritchard \& Smith }{1993}]{lps93}
Lyne~A.~G., Pritchard~R.~S., Smith~F.~G., 1993, MNRAS, 265, 1003

\bibitem[\protect\citename{Lyne {\rm et~al. }}{1996}]{lpsc96}
Lyne~A.~G., Pritchard~R.~S., Smith~F.~G., Camilo~F., 1996, Nat, 381, 497

\bibitem[\protect\citename{Manchester \& Peterson }{1989}]{mp89}
Manchester~R.~N., Peterson~B.~A., 1989, ApJ, 342, L23

\bibitem[\protect\citename{Manchester {\rm et~al. }}{1983}]{mnhg83}
Manchester~R.~N., Newton~L.~M., Hamilton~P.~A., Goss~W.~M., 1983, MNRAS, 202,
  269

\bibitem[\protect\citename{Melatos }{1997}]{mel97}
Melatos~A., 1997, MNRAS, 288, 1049

\bibitem[\protect\citename{Nagase {\rm et~al. }}{1990}]{ndl+90}
Nagase~F., Deeter~J., Lewis~W., Dotani~T., Makino~F., Mitsuda~K., 1990, ApJ,
  351, L13

\bibitem[\protect\citename{Newton, Manchester \& Cooke }{1981}]{nmc81}
Newton~L.~M., Manchester~R.~N., Cooke~D.~J., 1981, MNRAS, 194, 841

\bibitem[\protect\citename{Shemar \& Lyne }{1996}]{sl96}
Shemar~S.~L., Lyne~A.~G., 1996, MNRAS, 282, 677

\bibitem[\protect\citename{Siegman, Manchester \& Durdin }{1993}]{smd93}
Siegman~B.~C., Manchester~R.~N., Durdin~J.~M., 1993, MNRAS, 262, 449

\end{thebibliography}
\end{document}